# Chemistry-informed Machine Learning Explains Calcium-binding Proteins' Fuzzy Shape for Communicating Changes in the Atomic States of Calcium Ions

Pengzhi Zhang,[1,*] Jules Nde,[2,*] Yossi Eliaz,[3,4] Nathaniel Jennings,[3] Piotr Cieplak,[5] Margaret. S. Cheung[2,6]

1. Center for Bioinformatics and Computational Biology, Houston Methodist Research Institute, Houston, TX, USA

2. Department of Physics, University of Washington, Seattle, WA, USA

3. Department of Physics, University of Houston, Houston, TX, USA

4. Computer Science Department, HIT Holon Institute of Technology, Holon, Israel

5. Sanford Burnham Prebys Medical Discovery Institute, La Jolla, CA, USA

6. Environmental Molecular Sciences Laboratory, Pacific Northwest National Laboratory, Richland, WA, USA

* These authors contributed equally

**Abstract**

Proteins' fuzziness are features for communicating changes in cell signaling instigated by binding with secondary messengers, such as calcium ions, associated with the coordination of muscle contraction, neurotransmitter release, and gene expression. Binding with the disordered parts of a protein, calcium ions must balance their charge states with the shape of calcium-binding proteins and their versatile pool of partners depending on the circumstances they transmit, but it is unclear whether the limited experimental data available can be used to train models to accurately predict the charges of calcium-binding protein variants. Here, we developed a chemistry-informed, machine-learning algorithm that implements a game theoretic approach to explain the output of a machine-learning model without the prerequisite of an excessively large database for high-performance prediction of atomic charges. We used the *ab initio* electronic structure data representing calcium ions and the structures of the disordered segments of calcium-binding peptides with surrounding water molecules to train several explainable models. Network theory was used to extract the topological features of atomic interactions in the structurally complex data dictated by the coordination chemistry of a calcium ion, a potent indicator of its charge state in protein. With our designs, we provided a framework of explainable machine learning model to annotate atomic charges of calcium ions in calcium-binding proteins with domain knowledge in response to the chemical changes in an environment based on the limited size of scientific data in a genome space.

## I. Introduction

Proteins' fuzzy structures are features for communicating in biological processes,[1-3] instigated by secondary messengers such as calcium ions ($Ca^{2+}$). The binding of $Ca^{2+}$ almost always induces local or global conformational changes to the protein due to the alteration of its atomic charge or electrostatic interaction. These conformational changes and the dynamics of calcium-binding proteins are intimately linked to the function of a calcium-binding protein, indicating that the surrounding environments influence its structural dynamics, thereby impacting the accessibility of its several calcium-binding loops.[2,4,5] The versatility of calcium-binding proteins, often characterized by disordered domains, underscores the need of unraveling the causal relationship in determining the atomic charge states of $Ca^{2+}$ and the corresponding protein configurations. *Ab initio* calculations have been employed for this purpose, showing that the determination of the atomic charge of $Ca^{2+}$ requires information on its coordination geometry in the binding motif of a calcium-binding protein.[6] However, this approach is not only labor intensive due to in part the complexity of the protein environment involving water molecules, but also computationally expensive, making it infeasible to explore a broad configurational space of calcium-binding proteins in a changing environment. Because determining the atomic charge states is resource-intensive, it is extremely valuable to develop a machine learning (ML) algorithm that quickly identifies and explains the molecular descriptors influencing the atomic charge states without the necessity of a large database particularly in a genome space.

Calcium is a versatile signaling ion that plays a pivotal role in regulating downstream signaling targets in numerous biological functions, including muscle contraction, neurotransmitter release, and gene expression.[2,4,5,7-10] Calcium ion performs these biological functions through its interactions with numerous calcium-binding proteins, including calmodulin (CaM), which serves as the main calcium sensor and effectors.[11] Upon calcium binding, these proteins undergo conformational changes, enabling them to perform specific biological functions.[12,13] Additionally, the conformational changes reciprocally alter the ion charge state, which deviates from the conventional divalent charges (or +2e) observed in the crystal structures of calcium-binding proteins, due to the binding affinity that is influenced by the local protein environment and the presence of other ions or molecules.[14,15] The conformational changes affect the charge state, as atomic interactions are primarily dictated by the coordination number of a divalent ion,[16-18] in comparison to the atomic charges or the polarization effects.[19] It is essential for comprehending the intricacies of the causal functions,[14] such as the reciprocity observed among $Ca^{2+}$, CaM, and the CaM-target proteins in the context of synaptic plasticity.[12]

Here, we used a "sandbox" system of $Ca^{2+}$ and calcium-binding loops from shape-shifting CaM[20] as a use case to investigate how $Ca^{2+}$ brings the loops to adapt into distinct geometrical shapes dynamically, illustrating the role of their chemistry in the interplay of electrostatic forces, conformational changes, and molecular recognition within the biologically complex milieu. The calcium-binding loop is an evolutionarily conserved helix-coil-helix EF-hand motif that chelates $Ca^{2+}$ with its oxygen-rich amino acids, such as aspartate and glutamate acids, into a three-dimensional global form.[7] One of the amino acids from the EF-hand motif coordinates the $Ca^{2+}$ through a water molecule. This water molecule complicates the shape of the EF-hand motif[14] as if it rolls like a ball bearing, transitioning the EF-hand motif into various coordination geometries,[21-23] from octahedral to pentagonal bipyramidal. Alternating through these distinctive geometries of a calcium binding loop, $Ca^{2+}$ interacts with a combinatorial set of conserved and non-conserved amino acids that affect the charge state of the $Ca^{2+}$.

Despite a proliferation of ML models that establish the quantitative relations between configurational space and atomic charge or between configuration and energy in materials,[24-29] investigations into divalent ions such as $Ca^{2+}$ in protein environments are limited. This is due to a lack of accumulative data that covers a broad parameter space in biomolecules or materials, describing the causal relations for model training at the interface of organic and inorganic compounds. Among the burgeoning landscape of ML applications in molecular biology,[28,30-33] the interpretability of model outputs remains a critical challenge because changes in the input features nonlinearly contribute to the output. This creates a growing tension between the accuracy and interpretability of model predictions.[34,35]

To bridge the technical gap, we developed an ML algorithm (**Fig. 1**) that implements a game theoretic approach to explain the output of an ML model with a limited training dataset of over 7,000 configurations in our *sandbox* system. This dataset was curated from a combined *ab initio* calculations[14] and the structures of calcium-binding peptides from molecular dynamics simulations guided by the chemical coordination of $Ca^{2+}$.[36,37] We deployed network theory to uncover topological features that best capture the molecular descriptors to the charge states of $Ca^{2+}$ within the chemically complex datasets. We evaluated the SHAP (SHapley Additive exPlanations) values of these explainable features,[34,38] a concept rooted in cooperative game theory, for increasing transparencies of individual features that affect the ML models (see *Supplementary* for further description). The chemistry-informed ML algorithm explains the causal relationship between the unique characteristics of $Ca^{2+}$ embedded in various protein environments and their charge states.

We found that the coordination number of a $Ca^{2+}$ contributes the most to the determination of the calcium atomic charge state in various protein environments. With several other additive features, we applied the explainable models to annotate the calcium charge states of $Ca^{2+}$ binding motifs for another dataset of over 400 curated experimentally derived calcium-binding loops (**Table S1** in *Supplementary information*). The model annotates the atomic charge state of calcium according to the unique molecular descriptors of a configuration. Although the principle of target selection in calcium-binding proteins remains unknown due to the astronomical possibilities for a shape-shifting calcium-binding protein to connect with a versatile pool of downstream partners, we harnessed an explainable, chemistry-informed ML algorithm, to annotate calcium's atomic charge state by unraveling the causal features from the coordination chemistry that influence the shape of the loop in diverse protein environments, giving rise to robust regulations in calcium signaling.

# Chemistry-informed machine learning algorithm

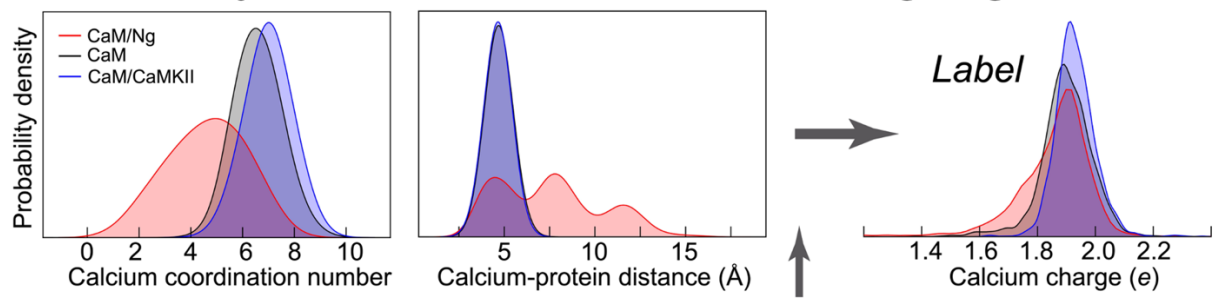

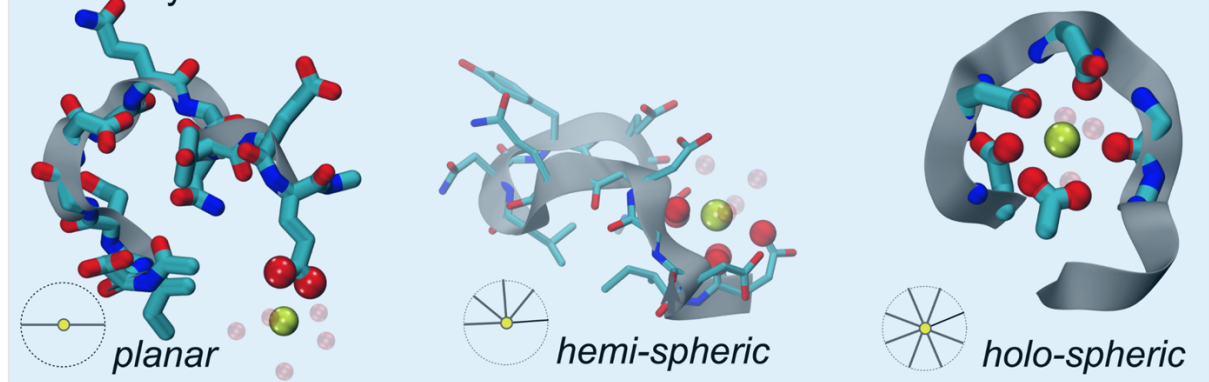

A variety of calcium environments extracted from MD simulations.

*planar*          *hemi-spheric*          *holo-spheric*

## Chemistry-informed feature extraction

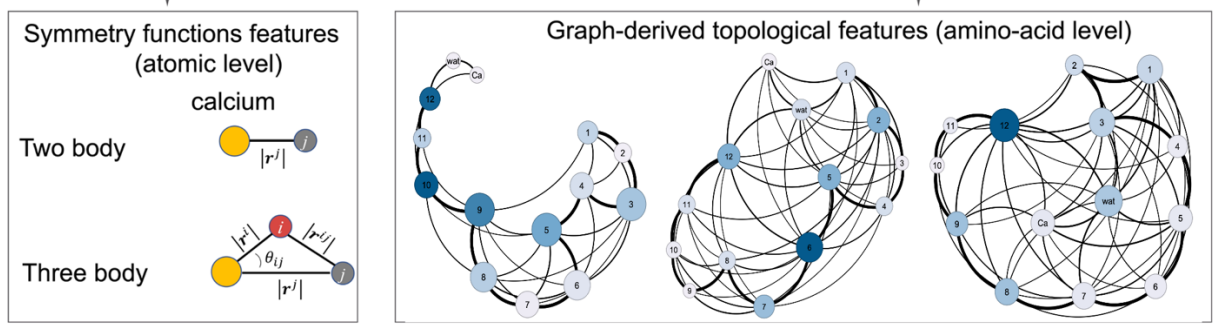

Symmetry functions features (atomic level)

calcium

Two body

Three body

Graph-derived topological features (amino-acid level)

## XGBoost predictive model

graph + symmetry features charge

Test

Training

### Explanation

Important features by ML model lead to better understanding of calcium binding in protein.

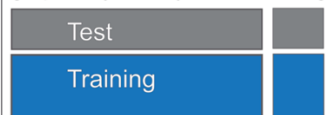

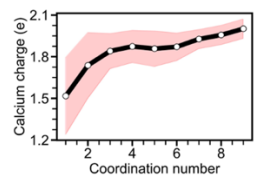

**Figure 1. Chemistry-informed ML enhances predictive power of atomic configuration-to-charge relation for Ca$^{2+}$.** A dataset of over 7000 configurations was curated from a combined *ab initio* calculations

of calcium-binding peptides and the structures of calmodulin-binding targets from molecular dynamics simulations guided by the chemical coordination of calcium [36,37] (the middle panel in light blue background). The shapes of the loops are categorized as planar, hemi-spheric, and holo-spheric coordination geometries. Variation of the loops shifts with the coordination numbers of calcium ions as well as the calcium-protein distances, tuning the atomic charge of the $Ca^{2+}$ (top panel). The graph-derived topological features were trained for explainable ML models to increase the transparencies and interpretability of features accounted by the many-body interactions. The additive features from many-body interactions further improved the accuracy of the ML algorithm compared to those that used only the two-body and three-body interactions as training features. The explainable XGBoost model computed and ranked the key features and relevant physical quantities in the charge state of calcium ions that impact most on the constructure of ML model.

## II. Results

## Overview: Chemistry-informed ML algorithm enhances the predictive power of a configuration-to-charge analysis by computing graph-derived topological features.

A single calcium ion is chelated by three to eight oxygen ligand atoms with an average of six ligands for all $Ca^{2+}$ binding loops or seven ligands for only EF-hand loops or by oxygen atoms from the nearby water molecules[37,39] (**Fig. 2a**). The chemical coordination of a calcium ion involves chelation with various classes of oxygen atoms, including those from the carboxyl groups (Asp, Glu), the carboxamide groups (Asn, Gln), and the hydroxyl groups (Ser, Thr) in the sidechains, as well as the carbonyl oxygen atoms of most residues in the main chain or cofactors, and water molecules, predominantly through electrostatic interactions. These oxygen atoms, in turn, are covalently linked to other heavy atoms such as carbon and nitrogen, forming a secondary shell. Additionally, a tertiary shell of carbon or nitrogen atoms can be delineated, consisting of atoms covalently bonded to the secondary shell.[37,39,40]

The configurations of ligands' oxygen atoms can be roughly classified into three categories based on their geometries: planar, hemi-directed (or hemi-spheric), and holo-directed (or holo-spheric) (**Fig. 1**). In a typical energy-minimized calcium-binding loop where the $Ca^{2+}$ is tightly bound holo-directedly, seven coordinating oxygen atoms are arranged to form a bipyramidal pentagonal coordination geometry, as shown in the crystal structure of $Ca^{2+}$/CaM (**Fig. 2a**). These concentric arrangements of atoms form a framework that defines the central binding site of calcium. Consequently, a set of physical parameters describing the spatial arrangement of atoms comprising the binding site can be delineated by the *"α_$r_c$"* / *"αβ_$r_c$"* for the radial/angular distribution functions. *"α"* or *"β"* is the element name (H, C, N, O) of atom surrounding the $Ca^{2+}$, and *"$r_c$"* is the cutoff distance within the concentric shell. The geometrical arrangement of the atom clusters also provides information about the closeness and betweenness centrality of $Ca^{2+}$, allowing for selective differentiation between $Ca^{2+}$ and other relevant divalent cations (see Supplementary **Fig. 1** for a complete description of features and the *Supplementary Information* for the definition of parameters).

We aim to use the curated conformations of the calcium-binding motifs in CaM and CaM-target bound complexes, which provide examples of the coordination geometries as a sandbox dataset, to train a chemistry-informed ML model. The domain knowledge of chemical principles (or multiple geometries of the calcium motifs) provided explainable molecular descriptors that influence the atomic charge state of calcium in various protein microenvironments. The proposed algorithm includes three major steps to capture these unique geometries from the structures:

*Step 1. Compute the feature of $Ca^{2+}$ and its coordination with a calcium-binding loop from the curated datasets:* With the sandbox dataset, we calculated the distribution functions to account for the spatial arrangement of the atoms surrounding the $Ca^{2+}$ and graph-based topological parameters to include high-order interactions. **Fig. 2b** shows pairwise interactions in terms of the contact probability map that captures probable interactions between the $Ca^{2+}$ and the conserved residues at the first, third, fifth, seventh, and twelfth positions, and the interaction between the ninth and twelfth residues. Featuring two-body interactions, we also capture the radial distribution functions around the $Ca^{2+}$. To capture the higher order interactions in the atom representation involved in the calcium coordination in a calcium-binding loop, we used a mathematical graph representation of the system and computed their topological parameters. The advantage of using measurements from graph theory over atomic contacts is that they capture the shape of the data in a complex system with voids and gaps in the structures.[41] The betweenness centrality of a node contains convoluted information about the entire graph of the calcium-binding loop, which may not be easily derived from a pairwise interaction map. For example, high betweenness centrality of the twelfth residue node suggests that the node is critical to maintain the connectivity of the network (**Fig. 2c**).

*Step 2. Establish explainable ML algorithms to evaluate additive features.* To assess the effectiveness of the extracted features, we designed three algorithms that learn the chemistry of many-body interactions in models. Model I: The radial distribution functions around the $Ca^{2+}$, featuring only the two-body pairwise interactions; Model II: In addition to Model I, the angular distribution functions around the $Ca^{2+}$, featuring the three-body interactions; and Model III: in addition to Model II, the topological order parameters from the graphs, featuring the higher order interactions. Then, we trained them with the XGBoost[42] regression models that were evaluated by the mean absolute error (MAE) in the calcium atomic charges, the coefficient of determination ($r^2$), and Pearson correlation coefficient (R) (see *Supplementary* **Figs. 2** and **3**).

*Step 3. Explain the feature importance for the interpretation of outcomes from the ML models.* The XGBoost regression model calculates the importance score according to the averaged contribution of a specific feature to the relevant decision trees. We explained the importance of each feature by their SHAP values.[34] The SHAP value, widely used in game theory to evaluate the marginal contribution of each player in a cooperative multiplayer game (see *Supplementary* for the equation and further details),  provides a principled way to explain the predictions of nonlinear models common in ML.  Positive (negative) SHAP values indicate that the feature positively (negatively) contributes to the model's prediction; thus, a higher positive (or negative) SHAP value suggests that the feature has a stronger influence on increasing (or decreasing) the prediction for a specific instance. Zero SHAP values imply that the feature does not contribute to the prediction for the given instance.

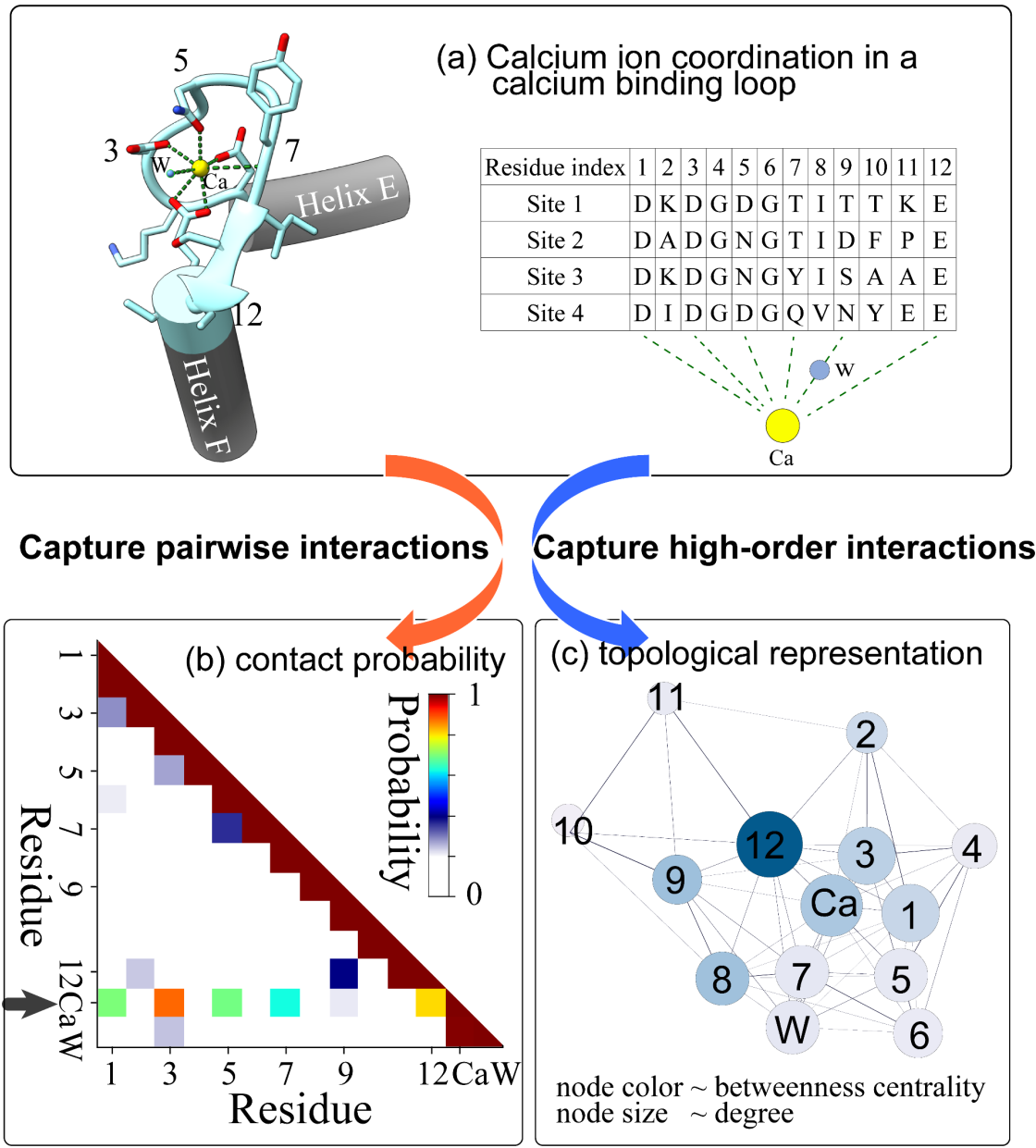

**Figure 2. Calcium-coordination features extracted to include pairwise and high-order interactions among residues in the calcium-binding loops.** (a) A schematic representation of the calcium-coordination structure. The calcium-binding loop is composed of 12 amino acids, and we relabeled the residue indices according to the amino-acid alignment. In this geometry, the coordination oxygen atoms come from the carbonyl groups in the sidechains of the glutamate/glutamine/aspartate/asparagine at the first, third, fifth, and twelfth position and the backbones of the amino acid at the seventh position (middle of the loop). The residues at the ninth position usually coordinate a Ca$^{2+}$ through a water molecule (W). (b) A contact probability map is

used for capturing the pairwise interactions between residues/ion/water molecules in the system. (c) A mathematical graph representation of the system and selected topological parameters captures the higher order interactions involved in the calcium coordination in a calcium-binding loop.

**Explainable machine learning reveals important additive features in training models**

To explain the importance of additive features, we constructed three explainable ML models while using the Pearson correlation coefficient (R) to test the accuracy of each model. We randomly divided the sandbox dataset into subsets for training (70%) and testing (30%), such as shown in Supplementary **Fig. 2a**. We evaluated the performance of the three explainable ML models with additive features in predicting charges in the test dataset **Fig. 3a**. When we used only the two-body pairwise interactions as features in Model I, the yielded ML model predicts the calcium atomic charge with the mean absolute error (MAE) ~0.07 e; there is a positive correlation between the predicted charges and the *ab initio* charges (R = 0.58). When the three-body interactions are also introduced as features in Model II, the accuracy of the ML model improved significantly, where R increases to 0.64. In the third explainable machine learning model in which the network topological parameters were included as features in Model III, there is a significant improvement of the accuracy on the test dataset (R increases to 0.71).

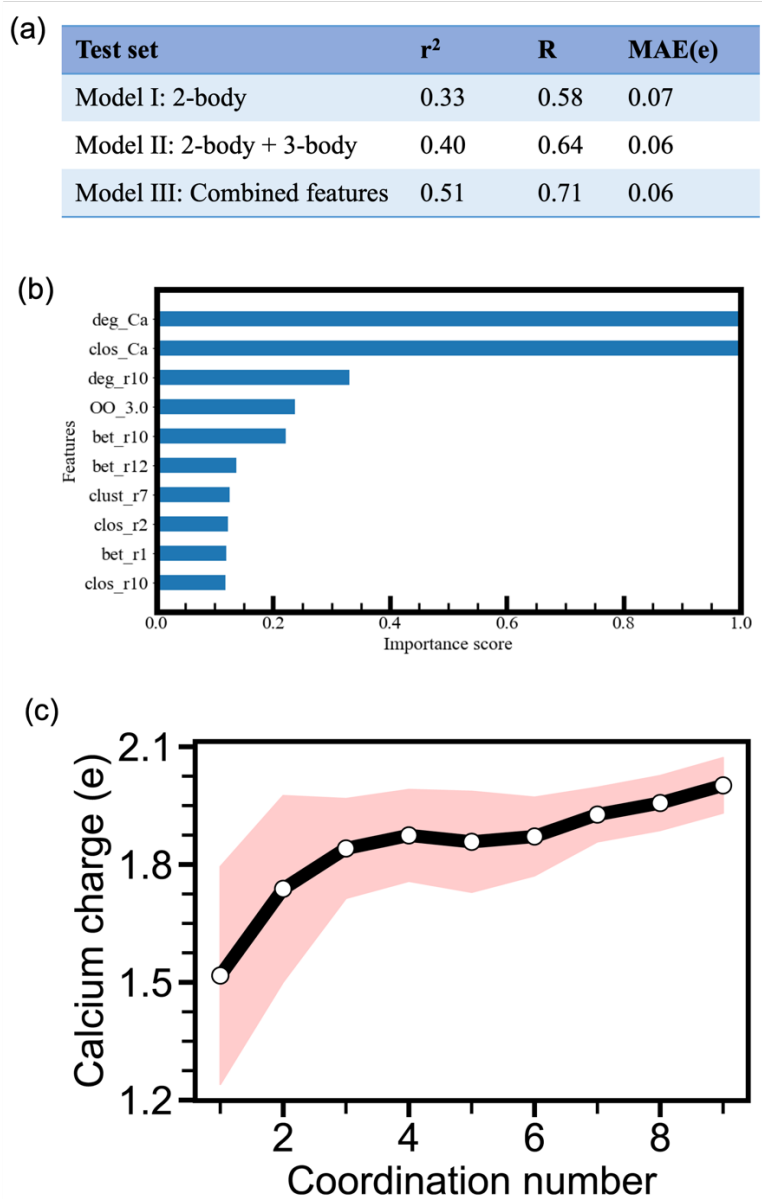

(a)

| Test set | r² | R | MAE(e) |
|---|---|---|---|
| Model I: 2-body | 0.33 | 0.58 | 0.07 |
| Model II: 2-body + 3-body | 0.40 | 0.64 | 0.06 |
| Model III: Combined features | 0.51 | 0.71 | 0.06 |

(b)

(c)

**Figure 3. ML performance and feature importance in determining the calcium charge**. (a) Evaluation statistics of models in predicting the calcium charge, $r^2$ is the coefficient of determination in the linear fit, R the Pearson correlation coefficient, and MAE mean absolute error. Addition of the network features noticeably improves the predicting performance in the test dataset when comparing Model I, II and III. (b) For Model III, we showed the ranking of the features that contribute most to the prediction of the calcium charge state; (c) for Model III, we showed the dependence of calcium charge on the coordination number, which reflects the top two most important features plotted against each other; mean and standard deviation (pink shaded area) of the calcium charge are plotted against coordination number of $Ca^{2+}$.

For the rest of analyses, we used only the Model III. We sorted the features by their importance scores (**Fig. 3b**) from the best performing explainable machine learning model with network order parameters as additive features. Among the top ten most important features, eight are network order parameters. The top two are closely related to the calcium-coordination number. The network order parameters about the $Ca^{2+}$ such as the closeness (i.e. *clos_Ca,*), which describes the average distance between the calcium ion and other coordination residues, and degree centrality (*deg_Ca*), which describes the total number of connections between the calcium ion and other coordination residues, appear most decisive to the charges of calcium ions. These two features show that calcium plays a central role in the loop by bringing in positive charges while other elements in the loop mostly contribute with negative charge. Surprisingly, the third most important feature is the network order parameter about the tenth residue node (i.e. *deg_r10*), a non-conserved residue in the calcium loops. This residue is not directly involved in the calcium coordination.

We then investigated the dependence of calcium charges on the coordination number (**Fig. 3c**), as the charges monotonically increases with the calcium coordination number. When the coordination number is less than 3, the mean of the calcium charges is less than 1.80 e and the standard deviation is large ~0.30 e. With such low coordination numbers, $Ca^{2+}$ is loosely bound in the loop forming a planar coordination geometry. There exist combinatorial ways to bind with oxygen candidates, contributing to the large deviation in the calcium charges. When the coordination number is between three and six, corresponding to a hemi-directed coordination geometry, the mean of the calcium charges is ~1.85 e and the standard deviation is ~0.15 e. When the coordination number is greater than or equal to six, the mean of the calcium charges monotonically increases with the coordination number from 1.87 e to 2.00 e with a small standard deviation ~0.07 e, forming a rigid holo-directed coordination geometry.

**SHAP values explain the contribution of additive features in annotating the calcium atomic charge states**

By interpreting a model trained on a set of features as a value function on a coalition of players in a theoretic game, SHAP values provide an intuitive way to evaluate the contribution of additive features to a prediction or to the uncertainty of a prediction. We aim to evaluate the contributions of the most influential descriptors in predicting the charge states not only globally (i.e. across the entire dataset), but also for the groups of high and low charges, which will further our functional understanding of the geometries of the calcium-binding loops, such as holo-directed and planar, observed experimentally. Therefore, we computed SHAP values on the entire dataset (**Fig. 4a**) and ranked the features based on their importance in predicting the calcium atomic charge states. The negative (or positive) Shapley values, indicating an effect of decreasing (or increasing) the calcium atomic charge state relative to the mean value. Low values of the deg_Ca and Clos_Ca variables have high negative contributions in predicting the calcium charge states, while high positive values have almost no contribution. These observations further solidify the central role of the calcium in the binding motif. We further computed the SHAP values of the structures of which the calcium charge is beyond 90 percentile (**Fig. 4b**), and the structures of which the calcium charge is below 10 percentile (**Fig. 4c**) to determine the overall importance of each feature in annotating the calcium atomic charge states in these subgroups. The top 90 percentile of the structures constitute a category of structures with "charges well above the mean value", while the bottom 10 percentile of the structures constitute a category of structures with "charges well below the mean value".

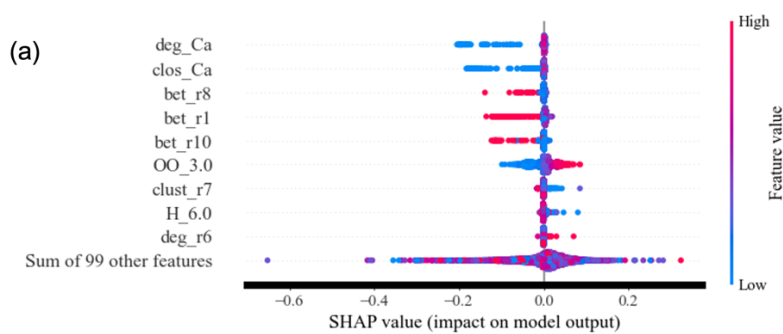

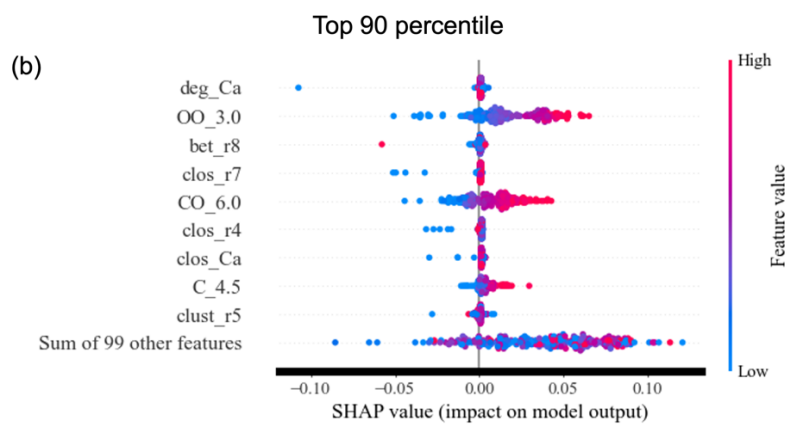

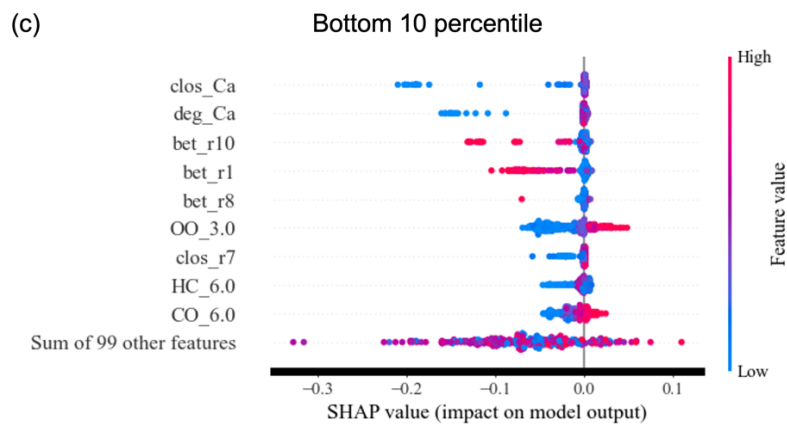

**Figure 4. Ranking of the features based on the SHAP values for their global contribution to the prediction of the calcium atomic charge state**. (a) The SHAP values of most important features influencing the calcium charge state prediction; (b) The SHAP values of the most important features influencing the charges that are above the 90th percentile; (c) The SHAP values of the most important features influencing the charges that are below the 10th percentile.

In examining **Fig. 4(a, c)**, we noticed a striking similarity among the most important features. The closeness centrality of calcium ion (how central is the calcium ion from the other atoms in the binding loop) together with its degree centrality (how connected the calcium ion is to other atoms in the binding loop) are the most dominant features. Both contribute to decreasing the calcium atomic charge state with lower feature values. This is expected, since calcium ion is the most valuable player in contributing its own charge state. Interestingly, the next top features are those derived from the topological graph network for residues 1, 8 and 10, whose removal in the game theoretic approach contributes to decreasing the calcium atomic charge state annotation. Similarly, in **Fig. 4b**, the topological order parameters for residues 7 and 8 contribute to decreasing the calcium atomic charge state annotation. These examples signify the importance of the geometry of the calcium binding loop to the determination of calcium charge states.

**Features that influence calcium charge states can be quite diverse for individual charge predictions**

In this section, we unraveled the local impact of each feature on the variability and patterns observed in calcium ion charge states. We calculated the SHAP values for each feature from Model

III (in **Fig. 4**) for a single instance, i.e., a specific calcium atomic charge state. The features that contribute the most to calcium atomic charge above an average value in the database are dominated by diverse features derived from the symmetry functions (**Fig. 5(a, c)**). These features mostly contribute to increasing the charges, except the $Ca^{2+}$ and carbon atoms interactions (carbon atoms from the third concentric shell) as well as the angular distribution between hydrogen, oxygen, and $Ca^{2+}$, which both contribute to decreasing the value of the annotated charge (**Fig. 5c**). On the other hand, the features that contribute the most to the charges below an average value (**Fig. 5(b, d)** are dominated by diverse features derived from the topological order parameters. These selections of the topological order parameters that contribute to decreasing the calcium atomic charge states, however, are quite diverse. These analyses signifying that it's the overall shape of the calcium-binding loop, rather than the exact residues, determining the value of calcium charges.

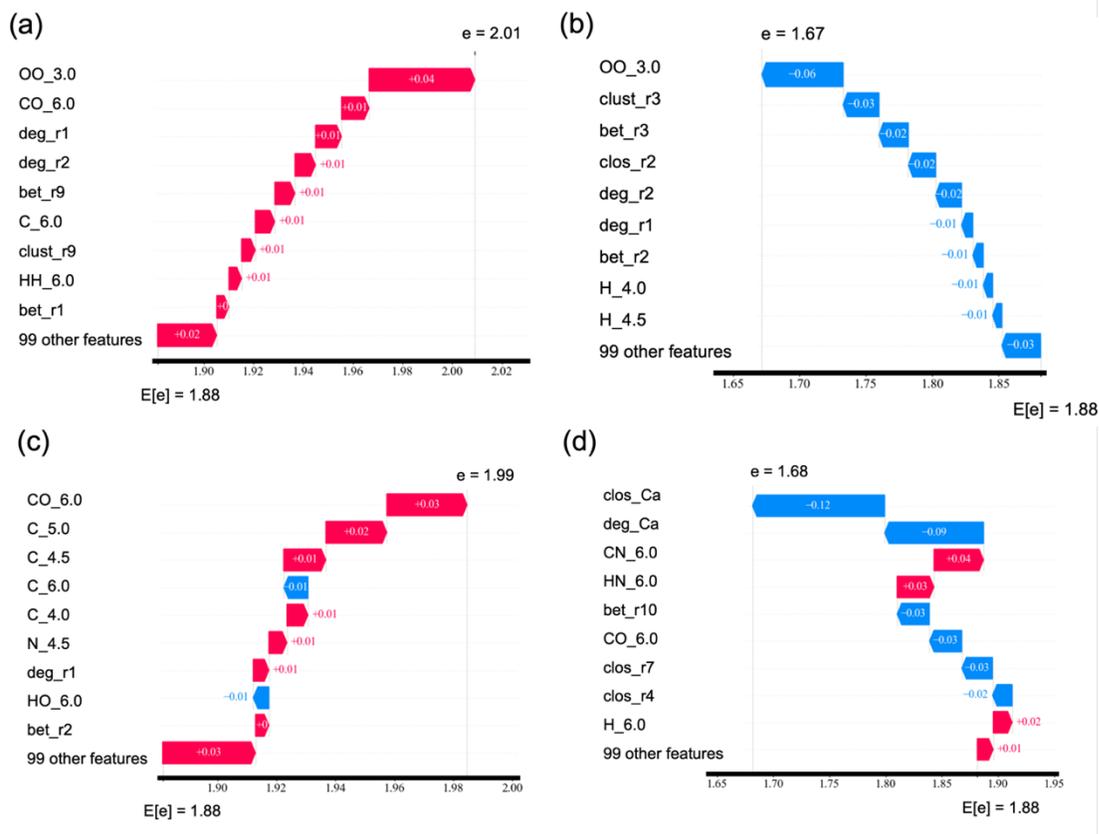

**Figure 5. Local contribution of the most importance features on the charge state of a single conformation**. (a, c) represent the structures whose calcium charges are above the averaged charge, (b, d) represent the structures whose calcium charges are below the averaged charge.

**Using the explainable chemistry-informed ML Model III to annotate the calcium charge states for the experimentally derived calcium-binding motifs**

Next, we used the chemistry-informed ML Model III to annotate the calcium charge states of the experimentally determined calcium-binding loops. We extracted 475 calcium-binding motifs from 217 protein structures downloaded from the Protein Data Bank (see *Supplementary* **Table 1** and *Supplementary* **Fig. 4**). Among these structures, five were solved using the nuclear magnetic resonance experimental technique and 212 were solved using the X-ray crystallography experimental technique. Each extracted motif has 12 residues plus $Ca^{2+}$ (see *Supplementary* information on how to extract EF-hand motifs from the Protein Data Bank structures) to match the data format of conformations in the *Sandbox* (https://github.com/Cheung-group/caXML). We provided the annotated charges of all the experimentally derived structures as *Supplementary* **Table 1** with the associated calcium motif structures.

We next evaluated the importance features that influence individual atomic charge state annotations. **Fig. 6(a, b)** show the cartoon representations of the calcium-binding loops annotated with an atomic charge above averaged value (**Fig. 6a**) and another one below the average value (**Fig. 6b**). (see *Supplementary* **Table 1** for the rest of the annotated calcium atomic charges). For the conformation in **Fig. 6a,** the geometry of the calcium binding loop is holo-spheric and its charge is close to 2.00 e. We used SHAP to access the most importance features influencing the charge in **Fig. 6c**. We observed the same trend in annotating individual calcium atomic charge state in **Fig. 5,** where the most influential features in a calcium binding are the symmetry functions capturing a holo-directed geometry. We analyzed the SHAP values for the other calcium binding loop whose calcium charge states is far below average in **Fig. 6d**. The shape of the calcium binding loop follows a planar geometry. The most influential features contributing to decrease of charges in this case are mostly topological order parameters. They exemplify the use of the geometry of a calcium binding loop to annotate its calcium charge state.

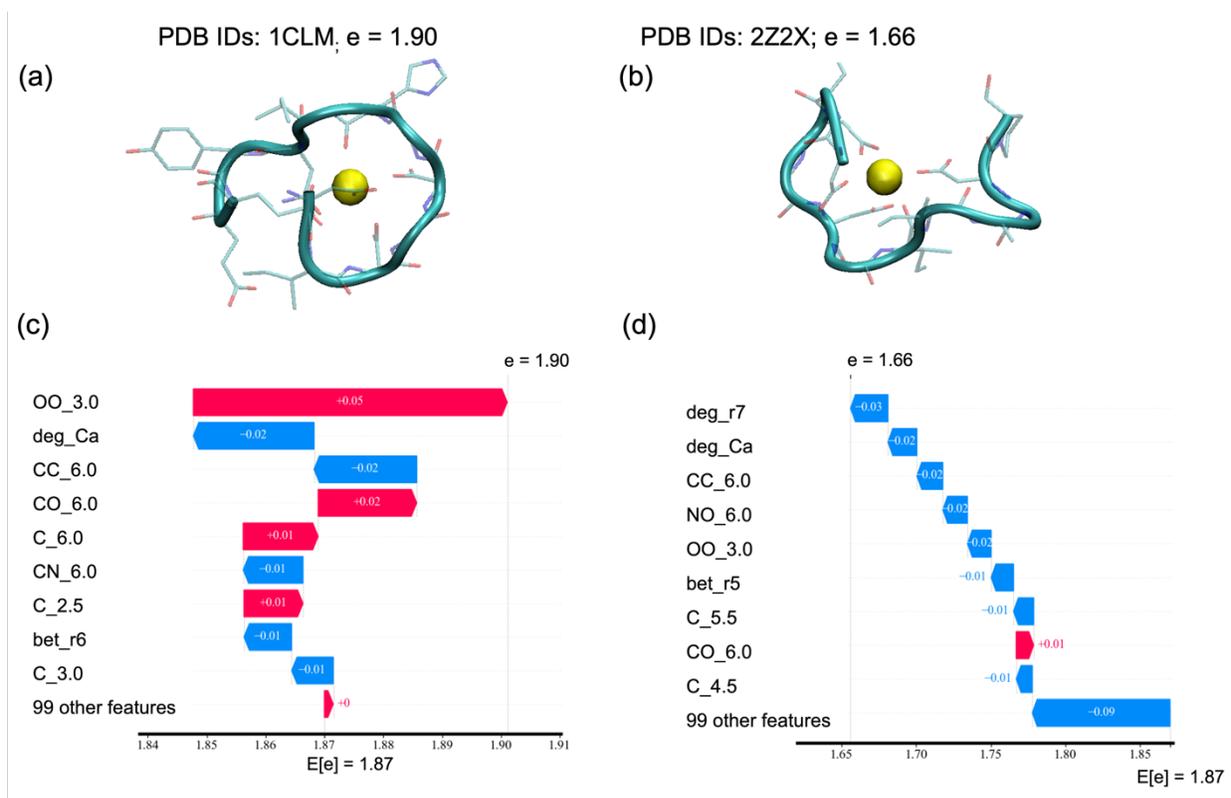

**Figure 6. Impact of input features on individual calcium atomic charge state annotation of the experimentally derived calcium-binding loops varies from charge to another.** Cartoon representations of the calcium-binding loops showing an annotated charge state (a) above or (b) below the average value. (c and d) The most influential features for representation in (a and b).

## III. Discussion

**Calcium-coordination number dictates the geometrical arrangement of the calcium-binding loop.**

Within calcium-binding loops, certain residues exhibit remarkable evolutionary conservation across diverse species. These conserved residues, typically found at positions one, three, five, seven, and twelve, play pivotal roles in coordinating $Ca^{2+}$ effectively[21,43,44] (**Fig. 2(a, c)**). Their preservation highlights their indispensable contribution to the structural integrity and functionality of calcium-dependent proteins. At position one, acidic residues such as glutamate (Glu) or aspartate (Asp) are prevalent, providing oxygen atoms that form strong electrostatic interactions with $Ca^{2+}$. Similarly, positions three, five, and seven often host residues with oxygen-containing functional groups, like Asp or serine (Ser), which enhance calcium coordination through hydrogen bonding.[43] Position twelve specifically features Glu with oxygen-rich side chain.[18,43] This residue at the twelfth position with a bidentate interaction contributes additional coordination sites, ensuring tight binding and specificity of the calcium-binding complex. These canonically conserved interactions are captured in the proposed chemistry-informed ML using the symmetry function features. The calcium binding loops coordinated by the five conserved residues adopt holo-spheric geometries (**Figs. 1** and **3**). From the cartoon representation shown in **Fig. 6a**, $Ca^{2+}$ is coordinated by Asp, Asp, Asp, His, and Glu at positions one, three, five, seven, and twelve respectively, which explains the holo-directed representation of such structure.

In contrast to the conserved core residues, calcium-binding loops often exhibit variability in other positions, where non-conserved amino acids are prevalent. These non-conserved residues introduce diversity into the binding loop, potentially modulating calcium binding specificity, affinity, or regulatory properties. Non-conserved amino acids may confer adaptability to environmental or physiological changes, allowing calcium-dependent proteins to adjust their function accordingly. Their presence could influence binding kinetics, coordination geometry, and overall structural dynamics of the binding loop, contributing to the versatility and functional diversity of calcium-binding proteins.[21] These non-conserved amino acids interact with $Ca^{2+}$ through the high order interactions, which are captured in the explainable chemistry-informed ML model using the topological graph features. In such representations, calcium binding loops are coordinated by at most three of the conserved residues. They may adopt either hemi-spheric or planar geometries (**Figs. 1** and **3**) depending on the number of coordinating ligands. From the cartoon representation shown in **Fig. 6b**, $Ca^{2+}$ is coordinated by Asp at positions one and three but is missing the conserved residue at positions five, seven, as well as Glu at the twelfth position. Therefore, the structure adopts either a hemi-directed or a planar geometry (**Fig. 1**). *The balance between evolutionary conservation and functional diversification not only maintains the essential functions of calcium-dependent proteins but also provides the flexibility necessary for adaptive responses to cellular cues and environmental stimuli.*

The relative mobility of the calcium reverses the intuitive physical effect of the chelating residues. In addition to the effect of mobile sidechains on the calcium charge, the loop backbone also presents a strong effect. For example, the oxygen atom from the ninth residue affects calcium through a mobile water molecule. The mobility of the water molecules contribute variations to the calcium local environment.. Similarly, the high mobility of water renders its contact with calcium almost invariant, so it cannot affect the charge. Residues at positions three and twelve, with the highest probability of contact (**Fig. 2b**), hold the calcium between them, and those residues determine the local environment of the calcium.

Since the twelfth residue in the loop, Glu, contributes to selectivity and specificity of calcium binding in proteins with EF-hand motifs, extensive studies have been conducted to understand its involvement in the structural stability of the EF-hand motif as well as the calcium charge state.[1,43,45] However, unlike the conserved twelfth residue, the tenth non-conserved residue has received less attention. The explainable chemistry-informed ML model showed that the tenth residue plays a topological role in maintaining the motif stability due to its position in the loop (**Fig. 3b**). Because this tenth position can be occupied by any other residue, mostly Phe, Val, Glu, Ala, Leu, Tyr, or Lys such as shown by Marsden et al.,[43] the contribution of this residue to the calcium atomic charge is determined by its topological importance rather than its chemistry.

**Topological network order parameters reveal hidden geometrical arrangement of residues in calcium-binding loops.**

It is advantageous to use network theory order parameter over atomic contacts or radial/angular distribution (i.e. symmetry functions) for capturing the hierarchical patterns in a complex data. In particular, betweenness centrality shows its importance as a critical bridge in connecting different nodes, while clustering coefficient indicates how tightly connected its immediate neighbors are. Topological network parameters play a critical role in distinguishing the calcium-binding loops from holo-spheric, planar or hemi-spheric geometries, such as evidenced in **Figs. 4c, 5(b, d),** and **6d**. These topological features signified the importance of the connectivity among $Ca^{2+}$ and the surrounding atoms, the closeness centrality of the $Ca^{2+}$, and the betweenness centrality of some of the non-conserved residues such as in the tenth position. With explainable features, we reveal the role of non-evolutionarily conserved residues in tuning the shape of a calcium-binding loop. The tenth residue is not directly involved in the coordination of $Ca^{2+}$ (**Fig. 2a**).[18] It is, however, critical in determining the geometrical arrangement of residues through the shape of the loop that can be modulated by CaM's interaction with its binding target.[11]

**SHAP explains the significance of additive features influential to the calcium atomic charge state.**

The explainable chemistry-informed ML represents a promising alternative to tackle the challenges presented by the *ab initio* representation of electronic structure data. One of the advantages of the chemistry-informed ML model over data-driven ML models (such as deep neural network) is that the features are tightly related to physical properties and thus are readily interpretable.[33,46,47] In the case of calcium atomic charge state, the algorithm learns from *ab initio* calculations data and the disordered fragments from the molecular dynamic simulations that contain information about $Ca^{2+}$ interactions within the different calcium-binding loops in the lobes of CaM and CaM/target complexes (**Fig. 2**). Through the recognition of patterns and relationships in the data, the model can be constructed to annotate the charge states of $Ca^{2+}$ based on features such as coordination chemistry and the different geometries of the calcium-binding loops accounted by the topological network parameters (**Fig. 1**).

In this ever-expanding landscape of ML, the interpretability of models remains a paramount concern.[48,49] In this context, the application of SHAP[34] emerges as a transformative tool for shedding light on the opaque inner workings of the algorithm (see *Supplementary* for further details). By decomposing complex predictions into understandable components, SHAP unravels the intricate relationships between additive input features and model outputs. Adding features from high-order interactions (topological network features) and from the symmetry function (radial/angular distribution functions) allows us to capture the loosely bound calcium geometry (planar or hemi-spheric) and the tightly bound calcium geometry (holo-spheric),

One of the most compelling aspects of SHAP lies in its ability to provide both global and local explanations[34,35] to understand how the combination of these features leads to an accurate calcium atomic charge state annotation. Global insights offer a holistic view of feature importance across an entire dataset (**Fig. 4(a-c)**), while local interpretations elucidate the rationale behind individual charge predictions (**Figs. 5(a-d)** and **6(c, d)**). This dual perspective enhances trust in the construction of ML model for charge annotation.

**IV. Conclusion**

The integration of chemistry-informed ML techniques explains the significant additive features in constructing models to annotate the atomic charge states of calcium ions in calcium-binding proteins, despite a small dataset in the genome or materials space. Combining the domain knowledge of chemistry, explainable ML, our approach identifies the explainable molecular descriptors in the fuzzy shape of a calcium binding loop that influence the calcium atomic charge state. It unveils the intricate patterns and correlations in additive features in a structurally complex dataset. With transparent machine learning models, subtle variations in the coordination chemistry of calcium ion in a calcium binding loop intricately modulate its charge states. This interdisciplinary approach not only enhances our comprehension of fundamental biological processes, but also offers invaluable insights into the design and engineering of metalloprotein-based systems with a limited genome space.

**V. Method**

The procedures of the explainable ML models are shown in the flowchart (**Fig. 7**), including the molecular dynamics data preparation, *ab initio* quantum-mechanics-based label generation, chemistry-informed feature extraction, experimentally derived calcium motifs data preparation, and construction and analyses of the ML model.

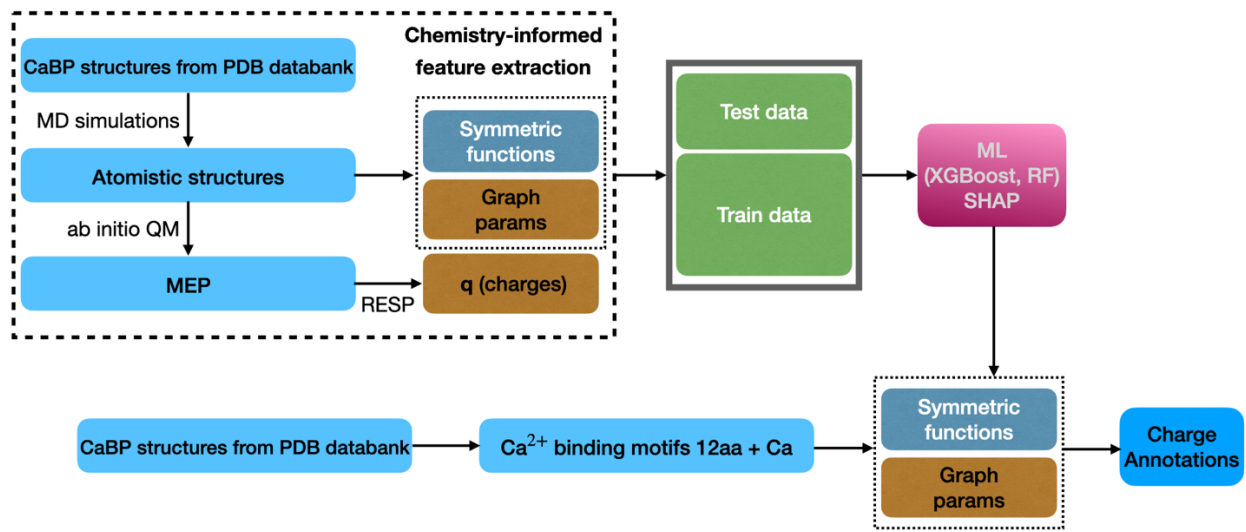

**Figure 7.** Flowchart of the chemistry-informed feature extraction and ML interpretation. The algorithm was trained using data from the combined *ab initio* electronic calculations and the structures of the calcium-binding loops from molecular dynamics simulations. We subsequently used one of the constructed ML models to annotate the calcium charge states from the calcium bind binding motifs extracted from the experimentally derived structures from the Protein Data Bank (Supplementary **Table 1** and Supplementary **Fig. 4** in the *Supplementary*).

**Data preparation: calcium-binding environments extraction and *ab initio* calcium charge determination.**

We consider a dataset consisting of selected calcium-binding loop structures with bound $Ca^{2+}$ ions and nearby water molecules from molecular dynamics simulations. Simulations of $Ca^{2+}$/CaM in explicit water in the following three conditions were examined: (i) Neat $Ca^{2+}$/CaM; (ii) $Ca^{2+}$-retaining environment: $Ca^{2+}$/CaM/CaMKII peptide; and (iii) $Ca^{2+}$-releasing environment: $Ca^{2+}$/CaM/Ng peptide. Please refer to the *Supplementary* for details of the molecular dynamics simulations. In total, 8,000 snapshots of $Ca^{2+}$ in EF-hand loops were selected for further analyses. Atomic charges were determined by fitting to the electrostatic potential obtained from single-point quantum mechanical calculations of the calcium-binding loop configurations. Detailed workflow of the charge determination is described in the *Supplementary* and in our previous study.[14]

**Topological parameters in the calcium-coordination network highlight many-body interactions at the amino-acid level.**

The topology of the calcium-coordination network was described analytically in the form of graph theory. The nodes are formed by an amino-acid, $Ca^{2+}$, or the closet water molecule. An edge is assigned when the distance between the closest pair of atoms of C, N, or O in each residue. For each residue in a loop, an undirected edge exists between the node representing it and its contacting residue with a weight equal to the reciprocal distance of the contact. Analysis of the graphs formed under this definition is accomplished in Gephi[50] and inherits its definitions of graph theoretic measurements.

Degree centrality, $C_d(v)$, quantifies how well the node connects with the rest of the graph. It is calculated as the degree or the connection number ($k$) of a node $v$ divided by the number of other nodes in the graph ($n$-$1$), such as shown in Equation (1)

$$C_d(v) = \frac{k}{n-1} \tag{1}$$

Betweenness centrality, $C_B(v)$, is used to describe how pivotal a node is to bridge the rest of the graph. It is calculated as the fraction that the node shortcuts any two other nodes in the graph. The mathematical definition of the betweenness centrality $C_B(v)$ of the node $v$ is defined in Equation (2)

$$C_B(v) = \sum_{\substack{s \neq v \\ s \neq t}}^{s \in V} \cdot \sum_{\substack{t \neq v}}^{t \in V} \cdot \frac{\sigma_{st}(v)}{\sigma_{st}} \tag{2}$$

with $\sigma_{st}$ representing the number of shortest paths from node $s$ to node $t$, and $\sigma_{st}(v)$ representing the number of those paths which go through the node $v$. The sum is over all pairs of nodes in the graph. V is the set of nodes.

Clustering coefficient, $CC(v)$, quantifies how well the neighbors of a node are interconnected. By definition, the clustering coefficient of a node $v$, is the fraction of edges to possible edges between its neighbors for the node $v$, as defined in Equation 3

$$CC(v) = \frac{T(v)}{\binom{k}{2}} \tag{3}$$

where $T(v)$ is the number of edges between the neighbors of the node $v$ and $k$ is the cardinality of the neighborhood (degree) of the node $v$.

Closeness centrality, $C_c(v)$, quantifies the average distance between the node and the rest of the graph. Mathematically, the definition is given in Equation 4

$$C_c(v) = \frac{n-1}{\sum_{u \neq v}^{u \in V} d(u,v)} \tag{4}$$

where d($u$,$v$) is the shortest path distance between nodes $u$ and $v$.

**Construction of predictive ML algorithms.**

We used a tree model with gradient boosting, XGBoost[42] and Sci-Kit Learn Python package,[51] to train and test the *ab initio* charge estimator models. We split the data into training (70%) and testing (30%) sets. Quality of the data splitting was justified by the overlap between distributions of the representative features and the label generated from the training/testing data (Supplementary **Figs. 2** and **3**). We used fivefold cross-validation by searching exhaustively on a hyperparameters grid on high-performance computers to determine the optimized hyperparameters. Specifically, we tuned *n_estimators* (number of gradient boosted decision trees),

*colsample_bytree* (the proportion of features for constructing each tree), *subsample* (the portion of training data samples for training each additional tree), and *learning_rate* (shrinkage of feature weights to make the boosting process more conservative and prevent overfitting, also called "$\eta$"). Subsequently, we made predictions on the calcium atomic charges using the optimized XGBoost regression model, assessed the model accuracy by calculating the coefficient of determination, Pearson correlation coefficient, and the MAE between the predicted and *ab initio*-derived reference values.

**Acknowledgment**

MSC and PZ thank the support from the National Institutes of Health (2R01GM097553). JN thanks the support from the National Science Foundation (MCB 2221824). We are grateful for the computing resources from XSEDE (TG-MCB190109), and the resources from the Research Computing Data Core at the University of Houston.

**Author contributions**

M.S.C., P.Z., Y.E., P.C. designed the research. P.Z., J.N., N.J. performed the research. M.S.C., P.Z., J.N., Y.E., P.C. analyzed data. All authors contributed to writing the paper.